# A Simple, Linear-Time Algorithm for x86 Jump Encoding

by Neil G. Dickson
http://www.neildickson.com/
December 25th, 2008

## Abstract

The problem of space-optimal jump encoding in the x86 instruction set, also known as branch displacement optimization, is described, and a linear-time algorithm is given that uses no complicated data structures, no recursion, and no randomization. The only assumption is that there are no array declarations whose size depends on the negative of the size of a section of code (Hyde 2006), which is reasonable for real code.

## The Problem

On x86 CPUs, there is more than one possible way to encode (assemble) a jump instruction, depending on how many bytes the jump is to traverse. For example, in 32-bit mode, an unconditional jump of 59 bytes can be encoded in 2 bytes of machine code (a short jump), but an unconditional jump of 237 bytes must be encoded as 5 bytes of machine code (a long jump). The problem that this causes is demonstrated in the following example:

```
LabelA:
    ...
    jmp     LabelB
    ...
    jmp     LabelA
    ...
LabelB:
```

In this case, the number of bytes required to encode the jump to LabelB affects the distance being jumped by the jump to LabelA, and vice versa. For example, if the jump to LabelB is a long jump, there are 3 more bytes between LabelA and the jump to LabelA than there would be if the jump to LabelB was a short jump. These 3 extra bytes adds to the distance jumped by the jump to LabelA, possibly meaning that the jump to LabelA now must also be a long jump, which then adds 3 bytes to the distance jumped by the jump to LabelB (and 3 bytes to its own distance from how backward jumps are encoded). The difficulty comes from that it is not initially known which jumps are short and which are long.

It should be noted that if it is known which jumps are short and which are long, it takes only linear time to encode them. On a first pass through, encode the jumps with the correct number of bytes but a distance of zero (or anything that will fit in the encoding). While doing this pass, record the byte offsets of labels in a hash by name (or unique ID), and record the byte offsets of jumps and the labels to which they jump in a list. Next, for each jump in the list, encode the distance jumped as the offset of the label it jumps to minus the offset of the jump, (minus the number of bytes to encode the jump, since jumps are relative to the following instruction). All of the jumps are then encoded correctly, because the offsets of the labels have not changed, and the offsets of the jumps have not changed. The problem then simply becomes to determine, in linear time, which jumps are short and which are long. This is important, since a linear number of jumps could each jump over a linear number of long jumps, meaning that updating each jump distance in-place whenever a jump is found to be long takes quadratic time in the worst case.

## The Algorithm

This algorithm relies on the fact that only up to 128 jump instructions, a constant number, can fit in 256 bytes. In practice, the average number of jump instructions per 256 bytes is on the order of 16. This is related to the allowed short jump distances.

| | Encoded Size (bytes) | Distance Range (bytes) |
|---|---|---|
| **Short Jump (any type)** | 2 | -128 to +127 |
| **Long Unconditional Jump in 32/64-bit mode** | 5 | -2,147,483,648 to +2,147,483,647 |
| **Long Conditional Jump in 32/64-bit mode** | 6 | -2,147,483,648 to +2,147,483,647 |
| **Long Unconditional Jump in 16-bit mode** | 3 | -32,768 to +32,767 |
| **Long Conditional Jump in 16-bit mode** | 4 | -32,768 to +32,767 |

The algorithm overall is as follows:

1. Encode all jumps as short jumps (with incorrect distances).
2. Calculate the actual distances currently spanned by all jumps.
3. Mark all jumps with distances outside of the short jump range, and add them to a queue, Q.
4. Until Q is empty,
    a. Remove jump, J, from Q.
    b. For each jump, K, within 128 bytes of J that are not marked and jump over J,
        i. Add (LongEncodedSize(J) – 2) to the absolute distance jumped by K.
        ii. If the distance jumped by K is now outside of the short jump range, mark K and add K to Q.
5. Re-encode, with marked jumps as long jumps and unmarked jumps as short jumps.

Each jump appears in Q at most once, and for each jump inserted into Q, at most 128 jumps are within 128 bytes of it (a 256-byte range). As such, at most $128n$ jumps are checked on line 4b, meaning that the whole loop through Q takes $O(n)$ time. Line 1, line 2, line 3, and line 5 each run in $\Theta(n)$ time, so the total time is $\Theta(n)$. Assuming that on average there are 16 jumps within 128 bytes of any long jump and that on average 1/8 of all jumps are long, only $2n$ jumps are checked on line 4b, meaning that the constant factor should be fairly low on average.

To see that the algorithm is correct, notice that beyond line 3, jump distances are tracked until they exceed the short jump range limits, and all short jumps that jump over J will be within 128 bytes of it, otherwise they wouldn't be short. After a jump has become long, no more increases of its absolute jump distance can affect any other jumps. Therefore, all changes that can affect whether a jump changes from short to long are taken into account, and so the marked jumps (those with a distance out of short jump range) are exactly those that are long jumps.

## More Specifically

There are a few subtle issues with the algorithm as stated above. For example, to check whether K jumps over J, the algorithm must use the offset and distance of K and J as encoded initially. This is because the relative positions of jumps and labels is the same with both the initial encoding and the final encoding, but in between, the distance tracked for jumps is not consistent with the offsets stored, since the offsets are not updated until the re-encoding.

Steps 1 and 2 must create an array of jumps, sorted by offset, where each jump has the following information stored:

| | Initial Value |
|---|---|
| **Offset when all short jumps** | (as found in step 1) |
| **Distance when all short jumps** | (as found in step 2) |
| **Current distance** | (copy of previous) |
| **Marked as long?** | false |
| **Size of long encoding** | 3, 4, 5, or 6 (depending on mode and type) |

Steps 3 and 4 then become:

- For i from 0 to n–1,
    - If Jump[i].currentDistance ≥ 128 or Jump[i].currentDistance < –128,
        - Jump[i].marked = true
        - Add i to Q
- While Q is not empty,
    - Dequeue j from Q
    - For i from j–1 decreasing while i ≥ 0 and (Jump[j].offset – Jump[i].offset) < 128
        - If Jump[i].marked = false and Jump[i].originalDistance > 0 and (Jump[i].offset+ Jump[i].originalDistance) > Jump[j].offset
            - Jump[i].currentDistance += (Jump[j].longEncodingSize – 2)
            - If Jump[i].currentDistance ≥ 128,
                - Jump[i].marked = true
                - Add i to Q
    - For i from j+1 increasing while i < n and (Jump[i].offset – Jump[j].offset) ≤ 128
        - If Jump[i].marked = false and Jump[i].originalDistance < 0 and (Jump[i].offset+ Jump[i].originalDistance) ≤ Jump[j].offset
            - Jump[i].currentDistance –= (Jump[j].longEncodingSize – 2)
            - If Jump[i].currentDistance < –128,
                - Jump[i].marked = true
                - Add i to Q

Steps 1, 2, and 5 are done as described earlier. However, re-encoding of non-jump instructions is not required except to fill in offsets that now have known values. Those encoded instructions may simply be copied into place as needed.

## Conclusion

The preceding algorithm is a deterministic, non-recursive, linear-time algorithm for encoding x86 jump instructions using only basic data structures (array, queue, hash) that is easily understood. This algorithm or a variation thereof will be used in the initial version of the assembler built into Inventor IDE.